\documentclass{article}

\usepackage{microtype}
\usepackage{graphicx}
\usepackage{subfigure}
\usepackage{booktabs} 

\usepackage{hyperref}

\usepackage[accepted]{icml2024}

\usepackage{amsmath}
\usepackage{amssymb}
\usepackage{mathtools}
\usepackage{amsthm}

\usepackage[capitalize,noabbrev]{cleveref}

\theoremstyle{plain}

\theoremstyle{definition}

\theoremstyle{remark}

\usepackage[textsize=tiny]{todonotes}

\icmltitlerunning{Are Protein Language Models Compute Optimal?}

\begin{document}

\twocolumn[
\icmltitle{Are Protein Language Models Compute Optimal?}



\icmlsetsymbol{equal}{*}

\begin{icmlauthorlist}
\icmlauthor{Yaiza Serrano}{nbd}
\icmlauthor{Alvaro Ciudad}{nbd}
\icmlauthor{Alexis Molina}{nbd}
\end{icmlauthorlist}

\icmlaffiliation{nbd}{Department of Artificial Intelligence, Nostrum Biodiscovery S.L., Barcelona, Spain}

\icmlcorrespondingauthor{Alexis Molina}{alexis.molina@nostrumbiodiscovery.com}

\icmlkeywords{Machine Learning, ICML}

\vskip 0.3in
]



\printAffiliationsAndNotice{}  

\begin{abstract}
While protein language models (pLMs) have transformed biological research, the scaling laws governing their improvement remain underexplored. By adapting methodologies from NLP scaling laws, we investigated the optimal ratio between model parameters and training tokens within a fixed compute budget. Our study reveals that pLM sizes scale sublinearly with compute budget, showing diminishing returns in performance as model size increases, and we identify a performance plateau in training loss comparable to the one found in relevant works in the field. Our findings suggest that widely-used pLMs might not be compute-optimal, indicating that larger models could achieve convergence more efficiently. Training a 35M model on a reduced token set, we attained perplexity results comparable to larger models like ESM-2 (15B) and xTrimoPGLM (100B) with a single dataset pass. This work paves the way towards more compute-efficient pLMs, democratizing their training and practical application in computational biology.
\end{abstract}

\section{Introduction}

Protein language models (pLMs) leverage natural language processing (NLP) techniques to model protein sequences, significantly transforming several scientific fields by enabling a deeper understanding of protein function and interactions, facilitating breakthroughs in drug discovery, protein design, and other biotechnological applications. 

Decoder-only pLMs excel in generative modeling, while encoder-only pLMs build robust representations that can be utilized for downstream applications. This dual capability enhances our ability to predict protein behavior and engineer novel proteins, driving innovation in computational biology.

Introduced in 2018, ESM-1 \citep{rives2021biological} signals the start of a transformative change in AI-driven biology, thanks to the transformer architectures \citep{vaswani2017attention}, that allowed scientists to solve complex biological challenges at unprecedented scales. These models, which varied in depth and parameters, facilitated the exploration of diverse protein sequence spaces such as UniRef100 \citep{suzek2015uniref}. 

Building on those results, the subsequent ESM-2 models \citep{lin2023evolutionary} incorporated subtle architectural changes and an increase in scale that led to improved performance, measured in several downstream tasks.

A plethora of models followed after these seminal works, that sought in scaling and different training strategies, the solution for increased performance. Prominent examples are the Ankh \citep{elnaggar2023ankh} and xTrimoPGLM \citep{chen2024xtrimopglm} models, which showed the benefits of increased compute and model size.

Parallel to these advancements, the relationship between model parameters, dataset size, and compute allocation became an increasing subject of study, as highlighted in \citet{hestness2017deep}. With the rise of large language models in NLP and their increasing parameter count, reference studies such as \citet{kaplan2020scaling} and \citet{hoffmann2022training} specifically addressed these scaling dynamics.

To the best of our knowledge, there has been no comprehensive effort focused in studying the relationship between parameter and token counts in the field of pLMs\footnote{During the review process of this work \citet{compute_optimal} was published.}. The closest to such work is \citet{hesslow2022rita}, and while they studied the effect of scaling in generative decoder-only models, they did not focus in what we consider has been the major success of pLMs, learning a representation of the underlying sequence space of proteins.

Therefore, we will focus on encoder-only models in this work, and we pose the following questions: \textit{In the context of pLMs, what is the optimal ratio between the number of tokens and model parameters given a fixed compute budget?} \textit{Are the widely used pLMs compute-optimal?} \textit{What is the optimal model size for a fixed compute budget?}

\section{Empirically obtaining compute-optimal protein language models.}

In this section, we will detail the adaptations done from \citet{hoffmann2022training} \footnote{In this work we omit \textit{Approach 2} from \citet{hoffmann2022training} due to computational constraints.} and \citet{kaplan2020scaling} leading to building empirical frameworks to assess the optimal ratio between training tokens and model parameters \footnote{Parameter counts for our models are defined at Appendix \ref{param:comp}.} given a fixed compute budget.

\textbf{Fix model sizes and vary the number of training tokens.} Mimicking \citet{hoffmann2022training}, we train several models ranging in size from 5M to 650M parameters (Appendix \ref{arch}) each with a different number of training steps. For each model size ($N$), we train for $s$ number of steps representing 4 different token sizes ($D$), ensuring that all tokens in the training set have been seen only once. 
Utilizing subsets of  UniRef50 protein sequence database, we adopted a simple sampling strategy to vary the number of training tokens. The strategy employed random sampling to generate several subsets with identical different token counts, ranging from 6.5B to 20.8B tokens.

For each model configuration and each run, we smoothed and interpolated the training loss curves. This smoothing process involved fitting a spline to the data points of FLOPs\footnote{FLOPs computation is equivalent to the one defined in \citet{hoffmann2022training} and can be found at Appendix \ref{flops:comp}.} and corresponding training losses. This allowed us to accurately interpolate the training loss for any given number of FLOPs.

Using the interpolated loss curves, we created a mapping from any FLOP count to the corresponding optimal model size and the number of training tokens. We evaluated the training loss at 1500 logarithmically spaced FLOP values for each run and identified the model size and token count that minimized the loss at each FLOP value. We then identified the most efficient models by selecting the model configuration that achieved the lowest loss at each FLOP value.

To derive the relationships \(N_{\text{opt}} \propto C^a\) and \(D_{\text{opt}} \propto C^b\), we fit power laws to the data of the most efficient models. Specifically, we used least squares to fit the model size and token count data as functions of FLOPs.The exponents \(a\) and \(b\) were determined from the fitted parameters.

\begin{figure}[ht]
\vskip 0.2in
\begin{center}
\centerline{\includegraphics[width=\columnwidth]{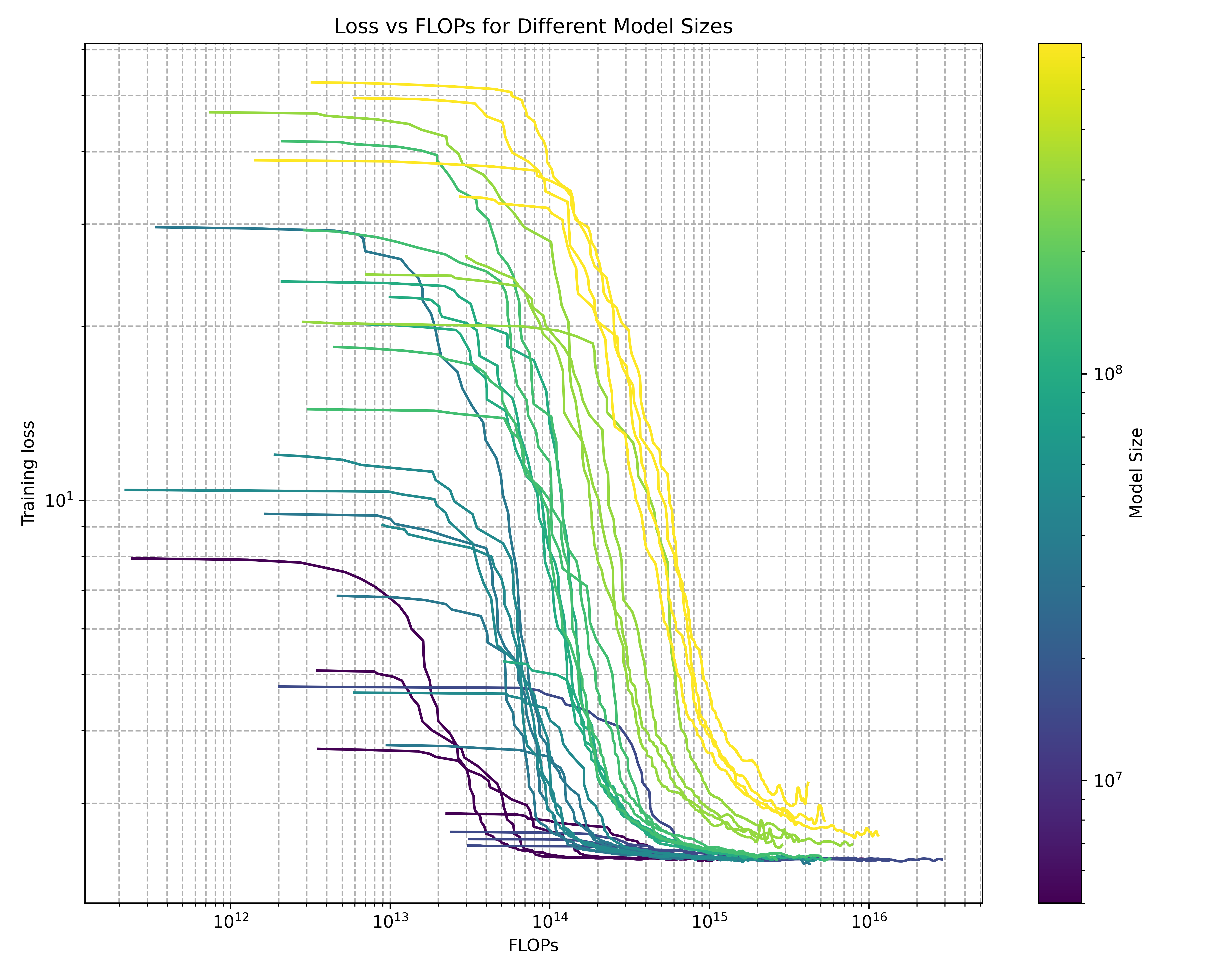}}
\caption{FLOPs vs loss with token-based learning rate decay.}
\label{loss_vs_flops}
\end{center}
\vskip -0.2in
\end{figure}
 
\textbf{Fitting a parametric loss function.} We analyzed the relationship between model size (parameters \( N \)) and dataset size (tokens \( D \)). The function used to fit the data was outlined in \citet{kaplan2020scaling}, which is further explain in Appendix \ref{scal:kap}. They obtained the combined scaling law:

\begin{equation} \label{equation_kaplan}
    L(N, D) = \left[\left(\frac{N_c}{N}\right)^{\frac{\alpha_N}{\alpha_D}} + \left(\frac{D_c}{D}\right)\right]^{\alpha_D}
\end{equation}

where, $N$ is the model size (number of model parameters). $D$ is the dataset size (number of tokens). $N_c$ and $D_c$ are critical values for model size and dataset size, respectively, and $\alpha_N$ and $\alpha_D$ are scaling exponents.

\section{Experiments and Discussion}

In this section we aim to solve the questions posed at the beginning of the manuscript through the proposed methods.

\subsection{What is the optimal ratio between the number of tokens and model parameters given a fixed compute budget?}

We sought to derive power laws that describe the relationship between the computational budget, measured in FLOPs, and the optimal model size and training tokens for a range of model configurations. With this purpose, we systematically vary the number of training steps across different model parameter sizes for a later fitting of power laws to the resulting training loss curves and a final fitting of the parametric loss with the data obtained. 

\textbf{Finding a power-law relationship between the parameter count, the number of tokens and computing budgets.}  By fitting power laws to the data of the most efficient models, we derived the relationships \(N_{\text{opt}} \propto C^{0.27}\) and \(D_{\text{opt}} \propto C^{0.71}\). These relationships suggest that the optimal model size scales sublinearly with compute budget, while the optimal number of training tokens scales superlinearly, indicating that encoder-only, protein language models exhibit diminishing returns in training loss improvements with increasing model size.

Intrigued by this findings we switched approaches to fit a parametric loss in order to obtain a more in depth explanation of the scaling parameters.

\begin{figure}[ht]
\vskip 0.2in
\begin{center}
\centerline{\includegraphics[width=\columnwidth]{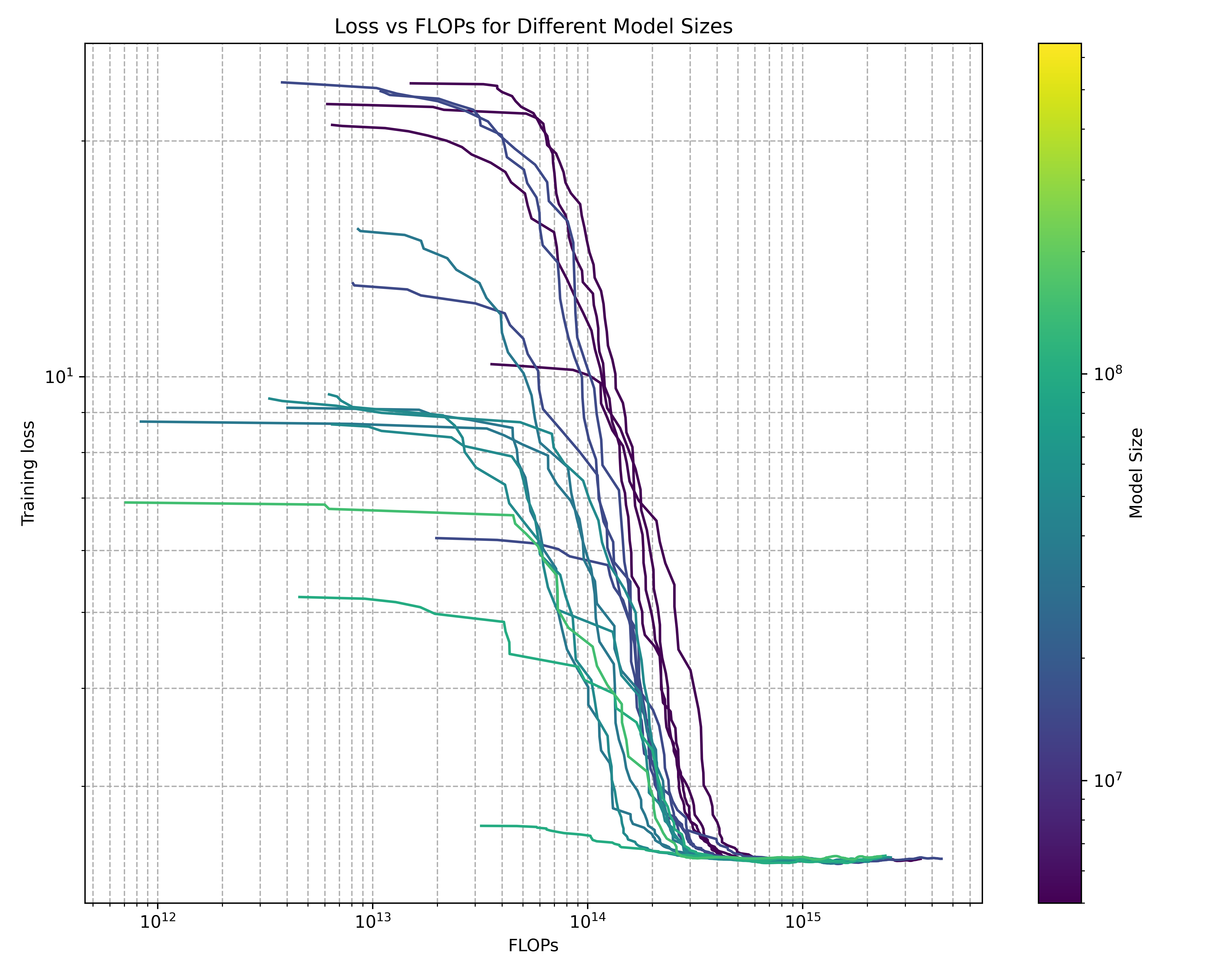}}
\caption{FLOPs vs loss without learning rate decay.}
\label{loss_vs_flops_2}
\end{center}
\vskip -0.2in
\end{figure}

\textbf{pLMs are lower-bounded in loss regardless of model size on a single dataset pass.} Using the described parametric loss in Equation \ref{equation_kaplan}, we used final losses after smoothing and interpolation. The small values of $\alpha_N$ and $\alpha_D$, reported in Table \ref{tab:fits}, indicate that the training loss for pLMs is relatively insensitive to changes in both model size $N$ and dataset size $D$. The extraordinarily large $N_c$ suggests that current model sizes are far from the critical point where size significantly impacts training loss. The unreasonable value of $D_c$ implies that dataset size does not affect the minimum achievable loss. These results demonstrate that increasing model size and dataset size offers minimal impact on reducing training loss within the explored range.

\begin{table}
    \centering
    \caption{Parameter fits to $L(N, D)$}
    \vskip 0.15in
    \resizebox{3.25in}{!}{
    \begin{tabular}{c|c|c|c|c}
        \toprule
        Parameter & $\alpha_N$ & $\alpha_D$ & $N_c$ & $D_c$ \\
        \midrule
        Value & 0.0063 & 0.0074 & $4.85 \times 10^{79}$ & $-1.33 \times 10^{69}$ \\
        \bottomrule
    \end{tabular}
    }
    \label{tab:fits}
\end{table}

Building upon the results adapted from \citet{hoffmann2022training} and \citet{kaplan2020scaling} we could conclude that none of the loss fittings performed was informative in terms of the impact that model size or training tokens can have in attaining a lower loss. It can be easily observed in both sets of training, i.e. with and without learning rate decay (see section \ref{x} and Figures \ref{loss_vs_flops} and \ref{loss_vs_flops_2}), that the learning dynamics of all models for all token sets are lower bounded by the same loss value (around 2.4\footnote{A similar plateau was detected in \citet{hesslow2022rita}, and similar loss values were found in \citet{chen2024xtrimopglm} and \citet{brandes2022proteinbert}.}).

\subsection{Finding the optimal tradeoff between parameters and model size to efficiently reach the observed plateau.} \label{x}

In order to avoid any forced dynamics in the convergence of the losses upon training, we re-fitted a subset of the models with the different token splits used in the previous section \footnote{Due to computational constraints not all models used for the approach explained before were trained.}. We speculated that the learning rate decay upon tokens seen in training could interfere with standard training processes in pLMs. We trained and evaluated models with the following configurations: 5M, 15M, 35M, 50M, 100M, and 150M parameters, covering a broad spectrum of model complexities. With this configurations, we observed that all of the fitted models reach the same plateau at the same loss value, regardless of the learning rate decay strategy. The realization that all training configurations reached a common plateau led us to question whether there was an alternative approach to optimize training for pre-trained language models (pLMs). Given a fixed number of FLOPs and tokens, we explored if this alternative could ensure optimal training when pLMs are trained on a single dataset pass, in relation to the observed plateau.

For each model size ($N$), we trained the models for a number of steps representing different token sizes ($D$). We focused on the loss and perplexity at the point where the models reached the plateau. We analyzed the relationship between model size, token count, and compute budget, by identifying the points where the models achieved different levels relative to the plateau. Specifically, we looked at 90\%, 95\%, and 100\% of the plateau value. For each level, we fit both linear and polynomial models to the data.
 
\begin{figure}[ht]
\begin{center}
\centerline{\includegraphics[width=\columnwidth]{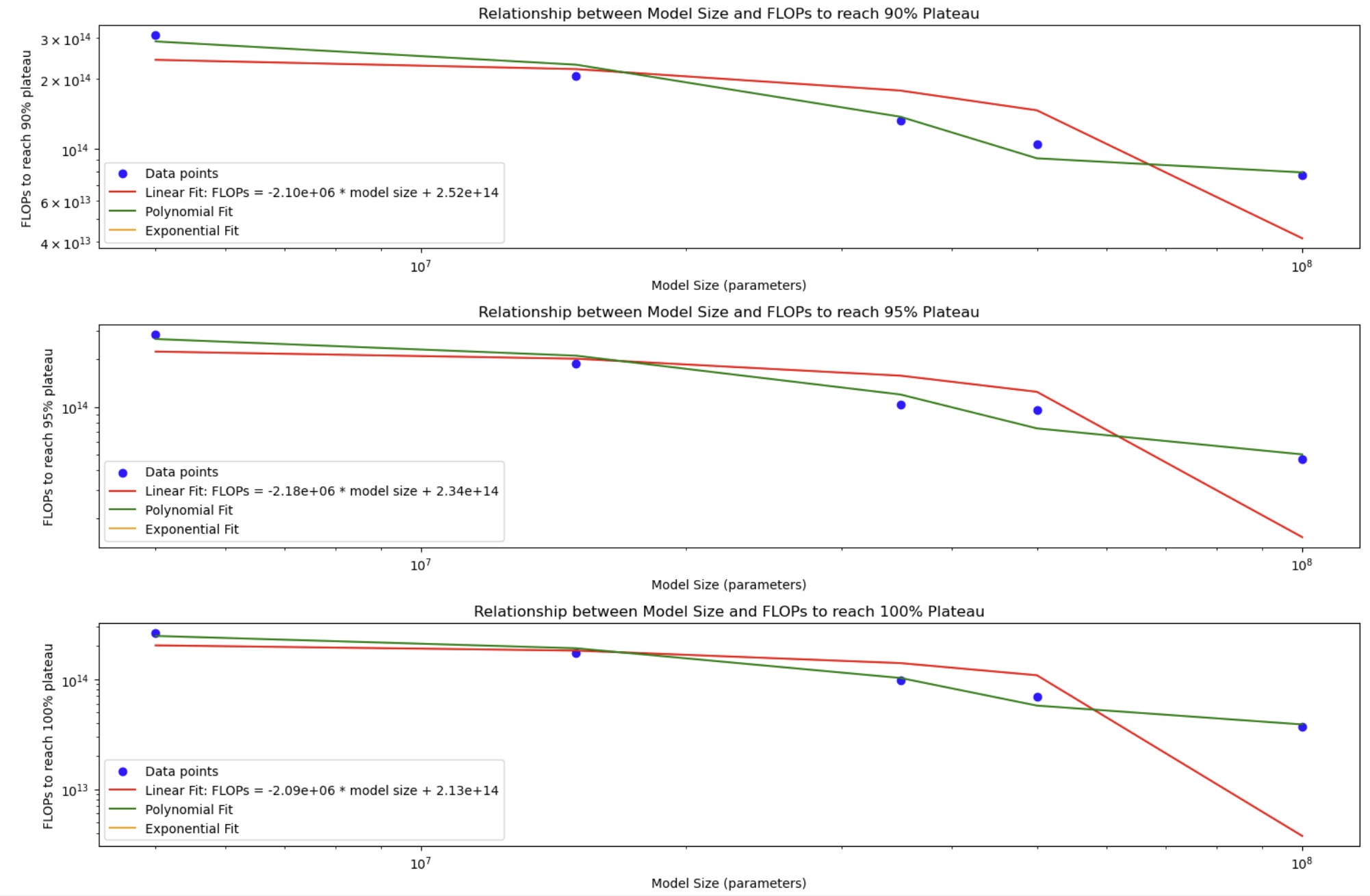}}
\caption{Fitting of lowest loss point per model size.}
\label{lowest_interpolation}
\end{center}
\vskip -0.2in
\end{figure}

The results from the linear regressions indicate a negative relationship between model size and the FLOPs required to reach various plateaus of the loss function. As compiled in Figure \ref{lowest_interpolation}, the negative slopes (\(-2.22 \times 10^{6}\) , \(-2.17 \times 10^{6}\) , \(-2.19 \times 10^{6}\)) suggest that as the model size increases, the FLOPs required to reach the plateau decrease. This might initially seem counter-intuitive because larger models typically require more compute. However, if larger models are more sample-efficient, they might reach the plateau loss more quickly despite their size. The large positive intercepts (e.g. \(3.04 \times 10^{14}\), \(2.61 \times 10^{14}\), \(2.38 \times 10^{14}\)) indicate the baseline FLOPs required to reach the plateau for smaller model sizes, reflecting the initial compute overhead. Larger models might have better inductive biases and higher capacity, allowing them to learn more efficiently from the same amount of data, thus reaching the plateau with fewer FLOPs.

\subsection{Are widely used pLMs trained following a compute optimal approach?}

The inability to achieve a significantly lower loss, regardless of training tokens and model size, suggests that scaling laws in the current form may not be fully applicable for pLMs. While loss reporting is sparse and perplexity comparisons across different pLMs are often not straightforward, we observed that this plateau is reached in several other works such as RITA, ProteinBERT, and xTrimoPGLM \citep{hesslow2022rita,brandes2022proteinbert,chen2024xtrimopglm}. It is evident that model size has little effect on this plateau, as demonstrated by the 1.2B parameter RITA model and the 100B parameter xTrimoPGLM model. Acknowledging that our model is trained in a single pass, we examined multiple-pass models such as ESM-2. In ESM-2, it is reported that between 270K and 500K training steps, there are diminishing returns, indicating that multiple passes over the same data have a minimal effect on improving performance.

To establish a fair comparison of performance, we trained a 35M parameter model on a 20M sequence subset of UniRef50 in a single pass until the model reached the loss plateau. We then tested this model on an 8K sequence set with a maximum of 0.5 identity with the training set. Our model achieved a perplexity of 11.43, compared to 10.81 reported for xTrimoPGLM and 10.98 for ESM2-15B \footnote{ESM-15B performance was extracted from \citet{chen2024xtrimopglm}}, a comparable value after taking into account the substantial reduction in compute, as we required far fewer FLOPs for training. For reference, in \citet{frey2024cramming}, the authors focused on efficient pLM pretraining, managing to create a foundational model with a reported perplexity of 13.72 in 24hrs while our model was trained for 1 hour in a single NVIDIA H100 GPU, achieving a marked improvement at a 11.43 of perplexity.

\subsection{Optimal model size to reach the observed plateau}

We used polynomial fitting to determine the optimal model size for a given compute budget, focusing on the relationship between model size and the FLOPs needed to reach various loss plateaus. This analysis predicts the model size that achieves the lowest loss efficiently within a fixed compute budget for a single dataset pass.

For a compute budget of \(10^{17}\) FLOPs, the optimal model size is approximately 50.6 billion parameters. For \(10^{18}\) FLOPs, it increases to 157 billion parameters. Extended results are compiled in Table \ref{table:optimal_model_sizes}.

These results show that as the compute budget increases, the optimal model size also increases. Larger models can utilize higher compute budgets more effectively to reach convergence plateaus. 

\begin{table}[htbp]
\centering
\caption{Optimal model sizes for different compute FLOPs budgets in a single dataset pass.}
\vskip 0.15in
\begin{tabular}{c|c}
\toprule
FLOPs & Optimal Model Size \\
\midrule
\(10^{17}\) & \(5.06 \times 10^{10}\) \\
\(10^{18}\) & \(1.57 \times 10^{11}\) \\
\(10^{19}\) & \(1.06 \times 10^{12}\) \\
\(10^{20}\) & \(7.35 \times 10^{12}\) \\
\bottomrule
\end{tabular}

\label{table:optimal_model_sizes}
\end{table}

\section{Conclusions and future work}

Our study suggest that NLP scaling laws are not transferable to pLMs. We found that regardless of model and dataset size, training reaches a plateau where no further improvements are observed. This points that the goal in training pLMs might be to reach this plateau as quickly as possible. Empirically, we observed that a single-pass training approach, stopped at the point of no improvement, achieves competitive results on held-out sets compared to larger models, trained with significantly more computational resources. 

While this work does not delve into the effects of multiple passes over the data during training, our comparisons with ESM-2 suggest that such approaches may lead to reduced perplexity when testing in samples close to the training distribution but with marginal gains in out-of-distribution samples.

In future work, we plan to extend this study with increased model and dataset sizes while evaluating downstream task performance. We also foresee further evaluation on the impact of different learning rates and multiple dataset passes. 
\bibliography{example_paper}
\bibliographystyle{icml2024}

\newpage
\appendix
\onecolumn
\section{Datasets}

We used UniRef50 \citep{suzek2015uniref}, retrieved in April 2024, which contains approximately 64 million non-redundant protein sequences grouped into clusters with at least 50\% sequence identity . Using the MMseqs2 algorithm \citep{steinegger2017mmseqs2}, we generated additional datasets: UniRef40, UniRef30, and UniRef20, where sequences share at least 40\%, 30\%, and 20\% sequence identity, respectively. Each cluster shows 80\% overlap with the seed sequence.

After obtaining these representative datasets, we created three additional datasets of the same sizes—Sampled30, Sampled23, and Sampled20—by randomly sampling from the original dataset. Table \ref{datasets} provides an overview of these datasets. For this study, we used only the second set of datasets due to computational limitations, reserving the first set for future experiments to assess dataset quality.

Additionally, we created a holdout set for the Sampled20 dataset, comprising 8K sequences with a maximum sequence identity of 0.5. These sequences were extracted from UniRef50, excluding those already included in the Sampled20 dataset.

\begin{table}[ht]
    \centering
    \caption{Overview of the datasets. Token amount is estimated from the number of sequences using the mean quantity of 325 aminoacids for sequence.}
    \vskip 0.15in
    \begin{tabular}{lccc}
        \toprule
        \textbf{Dataset} & \textbf{Sequences} & \textbf{Min seq id} & \textbf{Tokens} \\
        \midrule
        UniRef50  & 64M & 50\% & 20.8B\\
        \hline
        UniRef40  & 30M & 40\% & 9.75B\\
        UniRef30  & 23M & 30\% & 7.475B\\
        UniRef20  & 20M & 20\% & 6.5B\\
        \hline
        Sampled30 & 30M & 50\% & 9.75B\\
        Sampled23 & 23M & 50\% & 7.475B\\
        Sampled20 & 20M & 50\% & 6.5B\\
        \bottomrule
    \end{tabular}
    \label{datasets}
\end{table}

\section{Sequence embeddings}

We encoded protein sequences using 29 unique integer tokens: 20 for standard amino acids, 4 for IUPAC codes (\textit{B}, \textit{U}, \textit{O}, \textit{Z}), an unspecified amino acid (\textit{X}), a padding token (\textit{PAD}), a mask token (\textit{MASK}), and 2 special tokens (\textit{START} and \textit{END}). The \textit{START} and \textit{END} tokens were added at the beginning and end of each sequence to assist in processing sequences longer than 1024 amino acids. For sequences exceeding this length, we randomly selected a subsequence. Shorter sequences were padded with the \textit{PAD} token, and the \textit{MASK} token was used for masked language modeling.

Rather than fixing the number of sequences per batch, we fixed the number of tokens. Sequences were sorted by length and added to batches until the token limit was reached, with each batch padded to the length of its longest sequence.

\section{Architecture}\label{arch}
We employed the ESM-2 \citep{lin2023evolutionary} encoder-only protein language model, adjusting the model's layers, embedding size, and feed-forward hidden size for scalability. Table \ref{model_params} outlines eight models, varying from 5M to 650M parameters. Across all configurations, the number of attention heads remains constant at 20, and the feed-forward network's hidden dimension is consistently four times the embedding dimension, aligning with the original architecture.	
\begin{table}[ht]
    \centering

    \caption{Parameters of the ESM-2 models.}
    \vskip 0.15in
    \begin{tabular}{rccccc}
        \toprule
        Parameters & num\_layers & d\_model & ffw\_size & num\_heads & key\_size  \\
        \midrule
        5M   & 4  & 320  & 1280 & 20 & 16\\
        15M  & 8  & 400  & 1600 & 20 & 20\\
        35M  & 12 & 480  & 1920 & 20 & 24\\
        50M  & 15 & 520  & 2080 & 20 & 26\\
        100M & 23 & 600  & 2400 & 20 & 30\\
        150M & 30 & 640  & 2560 & 20 & 32\\
        300M & 32 & 880  & 3520 & 20 & 44\\
        650M & 33 & 1280 & 5120 & 20 & 64\\
        \bottomrule
    \end{tabular}
    \label{model_params}
\end{table}

\section{Training strategy}

The models were trained with a masked language modeling objective, using the BERT \citep{devlin2018bert} masking pattern where 15\% of residues in input sequences are corrupted and predicted. Of these, 80\% are replaced with \textit{MASK} tokens, 10\% with random amino acids, and 10\% remain unchanged. Cross-entropy loss was calculated only for the masked positions.

For optimization, we used Adam with $\beta_{1}=0.9$, $\beta_{2}=0.98$, $\epsilon=10^8$, and an L2 weight decay of 0.01. All models were trained for a single epoch on NVIDIA A100 GPUs. The number of devices used varied from 1 to 8, depending on the model size and availability. The last 35M model was trained on a single H100 GPU. Due to memory constraints, we did not fix the number of tokens per batch for all models, using a minimum of 5K and a maximum of 50K tokens per batch to optimize GPU memory usage.

Regarding the learning rate schedule, we explored two strategies: with and without learning rate decay. Inspired by \citet{hoffmann2022training}, we designed a token-based exponential decay that adjusts the learning rate based on the number of tokens processed during training, defined as:
\begin{equation}
    \text{lr} = \text{lr}_{\text{initial}} \times 0.999^{(\text{tokens} / 1,000,000)}
\end{equation}
This gradual reduction in the learning rate helps models converge at equivalent speeds. To contrast this, we also trained the models without applying learning rate decay. In both cases, we set the initial learning rate to $1 \times 10^{-5}$.

Training utilized DeepSpeed \citep{rasley2020deepspeed} for parallel computing and efficiency, with 32-bit precision and ZeRO stage 2 \citep{rajbhandari2020zero} to reduce memory usage. We optimized data processing with efficient loader worker processes and automatic memory pinning.

\section{FLOPS forward pass computation}\label{flops:comp}

We follow the protocol of \citet{hoffmann2022training} with a modification for taking into account the FLOPS assigned to the RoBERTa head used as a final layer. Embedding matrices are counted in both FLOPS and parameter counts, while non-linearities, biases and layer normalizations are omitted.
\begin{table}[ht]
    \centering
    \caption{Forward pass FLOPS computation. The backwards pass is assumed to have twice the amount of FLOPS as the forward pass.}
    \vskip 0.15in
    
    \begin{tabular}{c|c}
        \toprule
        Operation & FLOPs \\
        \midrule
         \textbf{Embeddings} & 2 x seq\_len x vocab\_size x d\_model \\
        \midrule
        \textbf{Attention Layer} & \\
        
        KQV projections & 2 x 3 x seq\_len x d\_model x (key\_size x num\_heads)\\
        Key @ Query & 2 x seq\_len x seq\_len x (key\_size x num\_heads) \\
        Softmax & 3 x num\_heads x seq\_len x seq\_len\\
        Softmax @ Query reductions &2 x seq\_len x seq\_len x (key\_size x num\_heads) \\
        Output projection & 2 x seq\_len x d\_model x (key\_size x num\_heads) \\
        \midrule
         \textbf{FFN Layer} & 2 x seq\_len x (d\_model x ffw\_size + d\_model x ffw\_size) \\
         \midrule
         \textbf{RoBERTa Head} & 2 x seq\_len x (d\_model x d\_model + d\_model x vocab\_size) \\
         \midrule
         \textbf{Total FLOPS} & Embeddings + num\_layers x (Attention Layer + FFN Layer) + RoBERTa Head\\
        \bottomrule
    \end{tabular}
    \label{tab:fits}
\end{table}

\section{Parameter count} \label{param:comp}
We follow the parameter count schema of \citet{kaplan2020scaling}, removing sub-leading
terms such layer normalizations and biases.
\begin{table}[ht]
    \centering
    \caption{Parameter computation}
    \vskip 0.15in
    
    \begin{tabular}{c|c}
        \toprule
        Operation & Parameters \\
        \midrule
         \textbf{Embeddings} & vocab\_size x d\_model \\
        \midrule
        \textbf{Attention Layer} & \\
        
        KQV projections & 3 x d\_model x (key\_size x num\_heads)\\
        Output projection & d\_model x (key\_size x num\_heads)) \\
        \midrule
         \textbf{FFN Layer} & 2 x (d\_model x ffw\_size) \\
         \midrule
         \textbf{RoBERTa Head} & (d\_model x d\_model) + (d\_model x vocab\_size) \\
         \midrule
         \textbf{Total Parameters} & Embeddings + num\_layers x (Attention Layer + FFN Layer) + RoBERTa Head\\
        \bottomrule
    \end{tabular}
    \label{tab:fits}
\end{table}

\newpage

\section{Scaling law from \citet{kaplan2020scaling}}

This approach is based on the following scaling laws. For models with a limited number of parameters, trained to convergence on sufficiently large datasets:

\label{scal:kap}

\begin{equation}
    L(N) = \left(\frac{N_c}{N}\right)^{\alpha_N}
\end{equation}

For large models trained with a limited dataset with early stopping:

\begin{equation}
    L(D) = \left(\frac{D_c}{D}\right)^{\alpha_D}
\end{equation}

When training with a limited amount of compute, a sufficiently large dataset, an optimally-sized model, and a sufficiently small batch size:

\begin{equation}
    L(C_{\min}) = \left(\frac{C_{c}}{C_{\min}}\right)^{\alpha_{C_{\min}}}
\end{equation}
With this laws, the combined scaling law is obtained as:

\begin{equation}
    L(N, D) = \left[\left(\frac{N_c}{N}\right)^{\frac{\alpha_N}{\alpha_D}} + \left(\frac{D_c}{D}\right)\right]^{\alpha_D}
\end{equation}
where, $N$ is the model size (number of model parameters). $D$ is the dataset size (number of tokens). $N_c$ and $D_c$ are critical values for model size and dataset size, respectively, and $\alpha_N$ and $\alpha_D$ are scaling exponents. This functional form is fitted using least squares.
\end{document}